\documentclass[aps,prl,twocolumn,showpacs,amsmath,amssymb,showpacs]{revtex4-1}
\usepackage{graphics,graphicx}
\usepackage{color}
\usepackage{simplewick}
\usepackage{amssymb,amsmath}
\usepackage{bm}
\usepackage{feynmp}

\begin{document}
\unitlength = 1mm
\def\vR{{\bf R}}
\def\vv{{\bf v}}
\newcommand{\hp}{{\hat{p}}}
\newcommand{\hk}{{\hat{k}}}
\newcommand{\vare}{\varepsilon}
\newcommand{\grad}{\mbox{\boldmath$\nabla_{\mbox{\tiny $\vR$}}$}}
\def\cY{{\cal Y}}
\def\ul#1{\underline{#1}} 
\def\sm#1{{\mbox{\tiny #1}}}
\newcommand{\be}{\begin{equation}}
\newcommand{\ee}{\end{equation}}

\graphicspath{
	{\string~/qccalcul/FFLO/surfaceFFLO/}
}

\title{Comment on ``Surface Pair-Density-Wave Superconducting and Superfluid States'' }

\author{Anton B. Vorontsov}
\affiliation{Department of Physics, Montana State University, Bozeman, Montana 59717, USA}
 
\date{\today}

\maketitle

\maketitle

The claim in \cite{Barkman2019} that pair-density wave (PDW) superconductivity in magnetic field is more stable near surface than in bulk,
is not supported by a microscopic theory. 

The Ginsburg-Landau (GL) analysis \cite{Barkman2019} is flawed in several respects: 
(i) it uses GL coefficients, Eqs. (2-4), that are valid only in the vicinity of
the tricritical point (TCP) where $H_{FFLO}(T) \approx H_0(T)$ \cite{Buzdin1997};
(ii) the truncated-gradient GL expansion is inadequate, since at high fields the modulation
wave-vectors $q$ are large and the gradient and magnetic energies are comparable $v_\sm{F} q/2 \approx \mu H $ \cite{Burkhardt1994,Matsuo1998} 
and have to be treated on the same footing;
(iii) properties of the surface region 
in the GL theory are not well-controlled:  
the bulk microscopic theory \cite{Buzdin1997} 
and the resulting GL functional miss the reduced symmetry of the interface \cite{Sigrist1991a}; 
moreover, boundary conditions for a fast-oscillating order parameter cannot be derived within GL theory, 
and require non-local microscopic treatment of quasiparticle scattering \cite{Ambegaokar1974,min99}.

I address these issues within the quasiclassical theory, 
that captures effects on lengthscales well below the 
pair coherence $\xi_0 = v_\sm{F}/2\pi T_c \gg k^{-1}_\sm{F}$, 
and is a tested tool for studying inhomogeneous FFLO states \cite{Burkhardt1994,Matsuo1998,VorontsovFFLO}. 
A singlet order parameter $\Delta(\vR)$ of symmetry $\cY_\hk$ (s-wave, d-wave, etc)
is determined from the self-consistency 
\be
\Delta(\vR) \langle |\cY_\hk|^2 \rangle \ln \frac{T}{T_c} = 
2\pi T \sum_{\vare_m>0} \left\langle \cY^*_\hk 
\left[ \frac{f_\uparrow + f_\downarrow}{2} - \frac{\Delta \cY_\hk }{\vare_m} \right]
\right\rangle_\hk
\label{sc}
\ee
where 
$\langle \dots \rangle_\hk$ denotes Fermi surface average. 
The instability is given by linearization of (\ref{sc}), 
corresponding to $\Delta^2$-terms in the free energy expansion (with arbitrary gradients).  
One integrates the linearised Eilenberger equations  (ODE) for the anomalous propagators
$f_{\uparrow,\downarrow}(\vR; \hk, \vare_m >0)$ 
\be
\left[ \frac12 \vv_\sm{F}(\hk) \cdot \grad + (\vare_m \pm i \mu H) \right] f_{\uparrow,\downarrow}(x,z; \hk, \vare_m) =
\Delta(x,z) \cY_\hk
\label{eil}
\ee
analytically, 
assuming an expansion 
\be
\Delta(x,z) = \sum_{m=0}^{n-1} \Delta_m \; \varphi_m(z) \; \eta(Q_x x)
\label{op}
\ee
in the orthonormal basis 
$\varphi_m(z) = e^{-z/2\xi_0} L_m(z/\xi_0)$ 
on $z\in[0,\infty)$, where $L_m$ are Laguerre polynomials. 
The cut-off, $n$, limits the extent of the order parameter along $z$ axis to $ \sim 4 n\, \xi_0$. 
The $x$-dependence $\eta(Q_x x) = \{ \exp(iQ_x x) \mbox{ or } \cos(Q_x x) \}$. 
Propagators evolve from the normal state $f_{\uparrow,\downarrow}(z\to\infty; \ul\hp)=0$,
to the $z=0$ interface where scattering can be modeled from specular reflection 
$f_\alpha(x,0; \hp, \vare_m) = f_\alpha(x,0; \ul\hp, \vare_m)$, 
to completely diffuse 
\cite{Nagato1996,*Higashitani:2015ic}. 
Substituting $f_{\uparrow,\downarrow}$ into (\ref{sc}) and projecting out $\varphi_m$-modes 
gives an eigenvalue problem that I solve to find the highest $H_c$ 
transition and the corresponding eigenvector $\Delta_{m=0 ... n-1}$.   

\begin{figure}[t]
\centering\includegraphics[width=0.90\linewidth]{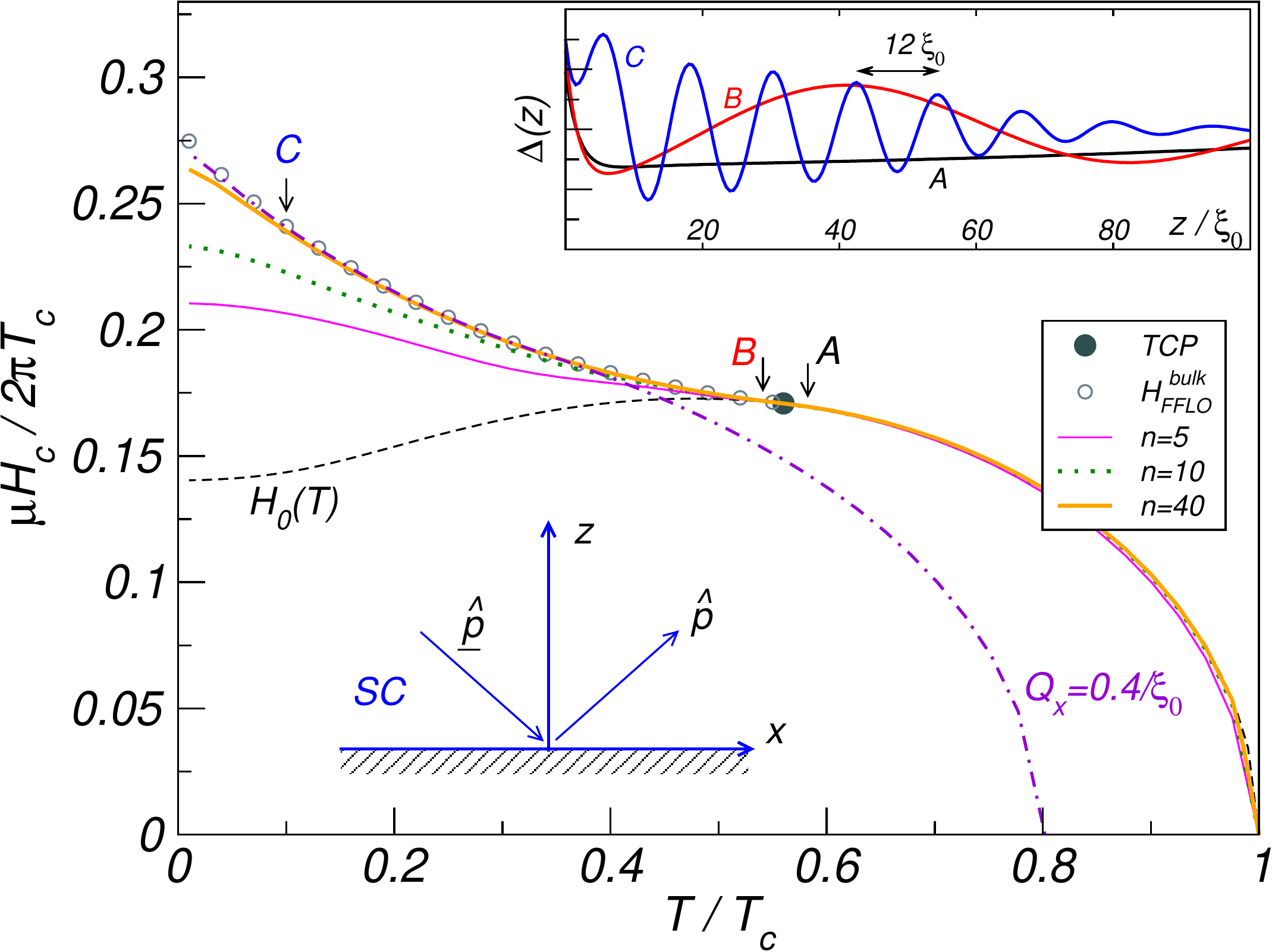}
\caption{
	The critical Zeeman field of a 2D s-wave superconductor with cylindrical Fermi surface, and atomically smooth
	boundary ($\ul\hp = \hp - 2\hat z(\hat z \cdot \hp)$). 
	$H_c$ for surface FFLO state (solid and dotted lines) 
	never exceeds the bulk critical field (open circles).  
	Inset: profiles of the order parameter emerging at transition points A,B,C. 
}
\label{fig}
\end{figure}

A superconducting state, localized within $20 \xi_0$ of the surface with $n=5$ components, 
is suppressed by $H_c$ field lower than the bulk critical field, see Fig.~\ref{fig}. 
As number of the basis components increases, the most favorable superconducting state expands away from the surface, 
and $H_c$ approaches $H^{bulk}_{FFLO}$. For $n=40$ we basically regain bulk behavior;
the superconducting state that appears at the transition  
changes from uniform everywhere above TCP, to modulated, with long periods just below TCP that become shorter as
temperature is lowered. 
At point C the period is $12 \xi_0$  - the value found in bulk system \cite{Burkhardt1994}. 

Solutions with finite modulation $Q_x$ \emph{along} the surface also never exceed the bulk transition. This
shows that the bulk instability is the most favorable one and there is no intrinsically more robust surface PDW
state in magnetic field.
\footnote{Enhanced surface critical field, found by the same authors in fully microscopic BdG calculations, 
has a close scaling relation with the higher zero-field $T_c$ in the surface atomic layer \cite{Samoilenka2020,*Samoilenka2020b}, 
indicating that FFLO physics is not the primary driving mechanism behind that enhancement. 
}
Similar results are also obtained 
for $d$-wave symmetry, for atomically rough (diffuse) boundaries, for 
Fermi surfaces of different shape (tight binding), in 1, 2, or 3 dimensions. 

\bibliographystyle{apsrev4-1}
\bibliography{mendeley,extra}

\end{document}